\documentstyle[12pt, aaspp4]{article}

\def\ltsima{$\; \buildrel < \over \sim \;$}
\def\simlt{\lower.5ex\hbox{\ltsima}}
\def\gtsima{$\; \buildrel > \over \sim \;$}
\def\simgt{\lower.5ex\hbox{\gtsima}}

\begin{document}

\title{Dynamics of a Massive Black Hole at the Center of a Dense 
Stellar System}

\author{Pinaki Chatterjee$^1$, Lars Hernquist$^2$ \& Abraham Loeb$^3$}
\affil{Harvard-Smithsonian Center for Astrophysics, 60 Garden
Street, Cambridge, MA 02138}
\footnotetext[1]
{pchatterjee@cfa.harvard.edu}
\footnotetext[2]
{lars@cfa.harvard.edu}
\footnotetext[3]
{aloeb@cfa.harvard.edu}

\medskip

\begin{abstract}

We develop a simple physical model to describe the dynamics of a massive
point-like object, such as a black hole, near the center of a dense stellar
system.  It is shown that the total force on this body can be separated
into two independent parts, one of which is the slowly varying influence of
the aggregate stellar system, and the other being the rapidly fluctuating
stochastic force due to discrete encounters with individual stars.  For the
particular example of a stellar system distributed according to a Plummer
model, it is shown that the motion of the black hole is then similar to
that of a Brownian particle in a harmonic potential, and we analyze its
dynamics using an approach akin to Langevin's solution of the Brownian
motion problem.  The equations are solved to obtain the average values,
time-autocorrelation functions, and probability distributions of the black
hole's position and velocity.  By comparing these results to N-body
simulations, we demonstrate that this model provides a very good
statistical description of the actual black hole dynamics. As an
application of our model, we use our results to derive a lower limit on the
mass of the black hole Sgr A* in the Galactic center.

\end{abstract}


\section{Introduction}

Black holes are thought to be ubiquitous in dense stellar systems.  Matter
accreting onto supermassive black holes near the centers of galaxies is
believed to be responsible for the energetic emission produced by active
galactic nuclei (Zel'dovich 1964; Salpeter 1964; Lynden-Bell 1969; Rees
1984).  Furthermore, it has been conjectured that all galaxies harbor such
black holes at their centers (but see Gebhardt et al. 2001 for recent
observations that some do not). Although definitive proof of this 
hypothesis is still lacking, there
exists evidence in some galaxies, such as NGC 4258 (Greenhill et al. 1995;
Kormendy \& Richstone 1995) and our own Galaxy (see Melia \& Falcke 2001
for a review), for the presence of an unresolved central dark mass of 
such high density that it is unlikely to be anything other than a black
hole (Maoz 1998). 
In the case of the Galactic center, which is thought to coincide with the
unusual radio source Sgr A*, future observations will measure the orbits of
individual stars within $0.1^{\prime\prime}$ of Sgr A* (see, e.g., Ghez et
al. 2000). In addition, forthcoming radio observations will significantly
improve the current limits on the proper motion of Sgr A* itself
(M. Reid 2001, private communication).  It is important, therefore, to
understand the general properties of the dynamics of massive bodies in
dense stellar systems so that the observations can be unambiguously
interpreted and predictions can be made to stringently test underlying
theories.

To pursue this goal, we present a simple model for the dynamics of a single
massive black hole at the center of a dense stellar system.  Our approach is
motivated by the recognition (Chandrasekhar 1943a) that the force acting on
an object in a stellar system broadly consists of two independent
contributions: one part, which originates from the ``smoothed-out'' average
distribution of matter in the stellar system, will vary slowly with
position and time; the second part, which arises from
discrete encounters with individual stars, will fluctuate much more
rapidly.

The smooth force itself is expected to be made up of two pieces: the
first is the force arising from the potential of the aggregate
distribution of stars at the position of the object; and the second is
the dissipative force known as dynamical friction, which causes the
object to decelerate as it moves through the stellar background
(Chandrasekhar 1943b).

The problem of the dynamics of a black hole in a stellar system is then
similar in spirit to the Langevin model of Brownian motion (see, e.g.,
Chandrasekhar 1943a), which describes the irregular motions suffered by
dust grains immersed in a gas.  In the Langevin analysis, a Brownian
particle experiences a decelerating force due to friction which is
proportional to its velocity, and it experiences an essentially random,
rapidly fluctuating force owing to the large rate of collisions it suffers
with the gas molecules in its neighborhood.

We extend this method of analysis to the black hole problem.  We take the
stellar system to be distributed according to a Plummer potential (see
Binney \& Tremaine 1987, hereafter BT) because the dynamical equations 
are then relatively tractable, and because this
density profile provides a reasonably good fit
to actual stellar systems.  In \S 2, we set up the model equations and also
provide a justification for breaking up the force on the black hole into
two independent parts, one smooth and slowly varying, and the other rapidly
fluctuating.  The equations of motion for the black hole are shown to be
similar to those of a Brownian particle in a harmonic
potential well.  In \S 3, we solve the equations of motion for the average
position and velocity of the black hole, and the time-autocorrelation
function of its position and velocity, obtaining both the transient and
steady-state components of these functions.  In \S 4, we derive the
probability distributions of the black hole's position and velocity by
solving the Fokker-Planck equation of the model.  It is shown that in the
steady state, these two variables are distributed independently with a
Gaussian distribution.  The conclusions of \S 3 and \S 4 are tested in \S 5
by comparing them with the results of N-body simulations of various
systems.  
In \S 6, we combine the results of the model and observational limits
on the proper motion of Sgr A* with physical arguments relating to the
maximum lifetime of the cluster of stars surrounding the black hole
(following the approach in Maoz 1998) to derive lower limits on the
mass of Sgr A*. Finally, \S 7 summarizes the paper.

\section{The Model}

Consider a black hole of mass $m$ in a cluster of stars which we take
to be described by a Plummer model of total mass $M$ and length
parameter $a$.  Thus, the density and potential profiles are given,
respectively, by 
\begin{equation} \label{dens}
\rho (r) = \frac{3 M a^2}{4 \pi} \frac{1}{(r^2+a^2)^{5/2}} ,
\end{equation}
\begin{equation} \label{pot}
\Phi (r) = -\frac{G M}{(r^2+a^2)^{1/2}} ,
\end{equation}
where $G$ is the gravitational constant and $\mathbf{r}$ is the radial
position vector from the center of the stellar system, which is taken
as the origin. The total mass in stars inside radius $r$ is then
\begin{equation} \label{mass}
M (r) = \frac{M r^3}{(r^2+a^2)^{3/2}} .
\end{equation}

Given the potential and density profiles, one can calculate the phase
space distribution function $f$, which in general depends both on
position $\mathbf{r}$, and velocity $\mathbf{v}$, and which is
defined such that $f({\bf r}, {\bf v})\,
d^3{\bf r}\,d^3{\bf v}$ is the mass in stars in the phase space
volume $d^3{\bf r}\, d^3{\bf v}$. We make the assumption that for the
spherically symmetric Plummer model, $f$ is a function of the relative 
energy
per unit mass ${\mathcal E}$ only (and independent of specific angular
momentum), where ${\mathcal E} = - \frac{1}{2}
v^2 - \Phi (r) = \Psi (r)  - \frac{1}{2} v^2$, $\Psi (r) = - \Phi
(r)$ being the relative potential. The distribution function can then
be calculated by the following equation (see BT):
\begin{displaymath}
f({\mathcal E})=\frac{1}{\sqrt{8} {\pi}^2} \int_{0}^{\mathcal E}
\frac{d^2 \rho}{d {\Psi}^2} \frac{d \Psi}{\sqrt{{\mathcal E} - \Psi}}.
\end{displaymath}
For the Plummer case, we get 
\begin{equation} \label{dist}
f({\mathcal E})=\frac{96}{7 \sqrt{8} {\pi}^3} \frac{M a^2}{(G M)^5} 
{\mathcal E}^{7/2}.
\end{equation}

With these preliminaries, we are in a position to calculate the forces
on the black hole in this model. There are three such forces: the
restoring force of the stellar potential, dynamical friction, and a
random force due to discrete encounters with stars.

The restoring force on the black hole of the stellar potential is
given by ${\bf F} = - m \nabla \Phi (r)$, where ${\bf r}$ is the
position vector of the black hole. Now
\begin{displaymath}
\Phi (r) = -\frac{G M}{(r^2+a^2)^{1/2}} = - \frac{G M}{a} \bigg(1- 
\frac{r^2}{2 a^2} + \frac{3 r^4}{8 a^4} + \cdots \bigg) .
\end{displaymath}
Since the black hole is much more massive than the stars, its typical
excursion from the center $r$ is small compared with $a$, and we are
entitled to neglect terms in the above equation of higher order than
$r^2$. The dominant restoring force on the black hole thus takes the form
of Hooke's law:
\begin{equation} \label{Hooke}
{\bf F} = - k {\bf r},
\end{equation}
where the ``spring constant'' $k$ is given by
\begin{equation} \label{sprconst}
k = \frac{G M m}{a^3}.
\end{equation}

As the black hole moves through the sea of stars, it experiences a
force of deceleration known as dynamical friction. We use for this the
Chandrasekhar dynamical friction formula (Chandrasekhar 1943b, BT):
\begin{displaymath}
{\bf F} = - \beta {\bf v},
\end{displaymath}
where
\begin{displaymath}
\beta=16 {\pi}^2 {\rm ln} \Lambda\, G^2 m (m+m_{\star}) \frac{\int_{0}^{v}
f(r,u) u^2 du}{v^3}.
\end{displaymath}
In the above, ${\bf v}$ is the velocity of the black hole, $m_{\star}$
is the mass of each star (in the following, we take all stars to have
equal masses, for simplicity), and ${\rm ln} \Lambda$ is the Coulomb
logarithm, which will be calculated below.

Note that the above formula was originally derived for the case of
a mass moving through a homogeneous stellar system, for which the
distribution function would be independent of $r$.  However, it is a good
approximation to replace this in the case of non-homogeneous systems
with the distribution function in the vicinity of the black hole (see
BT), especially since the distribution function for the Plummer model
varies slowly with $r$ in the region in which the black hole hole is
confined ($r \ll a$). Since the black hole moves very slowly compared
with the stars, we may replace $f(r,u)$ in the integral by $f(r,0)$ to
obtain (see BT):
\begin{displaymath}
\beta=\frac{16 {\pi}^2}{3} {\rm ln} \Lambda\, G^2 m (m+m_{\star}) f(r,0).
\end{displaymath}
But 
\begin{displaymath} 
f(r,u=0)=\frac{96}{7 \sqrt{8} {\pi}^3} \frac{M a^2}{(G M)^5} 
(- \Phi (r))^{7/2},
\end{displaymath}
and for $r \ll a$, $\Phi (r) \simeq - GM/a$. Thus, we finally get
\begin{displaymath} 
\beta=\frac{128 \sqrt{2}}{7 \pi} {\rm ln} \Lambda\, \bigg(\frac{G}{M
a^3}\bigg)^{1/2} m (m+m_{\star}),
\end{displaymath}
or, since $m \gg m_{\star}$,
\begin{equation} \label{b}
\beta=\frac{128 \sqrt{2}}{7 \pi} {\rm ln} \Lambda\, \bigg(\frac{G}{M
a^3}\bigg)^{1/2} m^2.
\end{equation}

The factor $\Lambda$ in the Coulomb logarithm is given by 
\begin{displaymath}
\Lambda = \frac{b_{max}}{b_{min}} \simeq
\frac{b_{max} V_{0}^{2}}{G m},
\end{displaymath}
where $b_{max}$ and $b_{min}$ are, respectively, the maximum and minimum
impact parameters between the black hole and the stars that need be
considered; $b_{min}$ is usually set to be (see BT, Maoz 1993) $b_{min}
\sim Gm/V_0^2$, where $V_0$ is the typical relative speed between the black
hole and the stars with which it interacts. Since the velocity of the black
hole is much smaller than that of the stars, we set $V_0^2$ to be the mean
squared velocity of the stars:
\begin{equation} \label{msvstar}
V_0^2=\overline{v^2(r)}=\frac{\int v^2\,f({\mathcal E})\, 4 \pi v^2\,
dv}{\int f({\mathcal E})\, 4 \pi v^2\, dv} = \Psi /2 \simeq \frac{GM}
{2a} \qquad \textrm{for small $r$}.
\end{equation}
The maximum impact parameter $b_{max}$ is not well-defined; however, an
error in the choice of $b_{max}$ results in a much smaller error in the
coefficient of dynamical friction, in which $b_{max}$ enters as the
argument of the logarithm function. (Note that there are other
uncertainties as well: in the above formula for dynamical friction, the
velocity dependence of the Coulomb logarithm was ignored by replacing it
with a typical relative velocity, $V_0$, which is treated as a constant;
when the velocity of the black hole is small, it is not clear that this is
a good approximation, and it is possible that the magnitude of the
coefficient of dynamical friction would be modestly reduced relative to the
above expressions [see Merritt 2001].)  In this paper, we adopt a
density-weighted formula for the Coulomb logarithm given by Maoz (1993),
which provides an implicit expression for $b_{max}$; in this case,
\begin{equation} \label{bmax} 
\int_{b_{min}}^{\infty} \frac{\rho (r)}{\rho (0)} \frac{dr}{r} ,
\end{equation}
should replace the Coulomb logarithm in the above equations. For the
case of the Plummer potential, we obtain  
\begin{equation} \label{lambda}
{\rm ln} \Lambda \rightarrow \sinh^{-1} \bigg( \frac{M}{2m} \bigg) - 
\frac{4}{3}
\frac{ \lbrack 1+3 (m/M)^2 \rbrack}{\lbrack 1+4 (m/M)^2 \rbrack^{3/2}}.
\end{equation}
This gives a value for $b_{max}$ that is somewhat smaller than the core
radius $a$ (a value often chosen for $b_{max}$), and indeed we find this
choice provides a slightly better fit to our simulations (see \S 5) than
the alternative choice of $b_{max}\approx a$.

The third force acting on the black hole is the stochastic force denoted as
${\bf F}(t)$, which arises from random discrete encounters between the black
hole and the stars.  This force cannot be written down analytically in
closed form, and is only defined statistically as described below.

We can therefore characterize the dynamics of the black hole by
\begin{equation} \label{veceqofmotion}
m \ddot{{\bf r}}(t) + \beta \dot{{\bf r}}(t) + k {\bf r}(t) = {\bf F}(t),
\end{equation}
which is the equation of motion of a harmonically bound Brownian
particle.
The spatial components of this linear vector equation are separable 
into equivalent terms, and we will without loss
of generality concern ourselves only with its $x$-component:
\begin{equation} \label{eqofmotion}
m \ddot{x}(t) + \beta \dot{x}(t) + k x(t) = F_x(t).
\end{equation}
(We will use $\dot{x}$ and $v_x$ interchangeably in the following.)  This
is a stochastic differential equation since, as noted above, the form of
$F_x(t)$ is not known. However, since this stochastic force is random and
rapidly varying, we expect: (1) that this force is independent of $x$; (2)
that it is zero on average; and (3) that to an excellent approximation this
force is uncorrelated with itself at different times. We may formalize
these statements as
\begin{equation} \label{stochdef}
\langle F_x(t) \rangle = 0, \qquad
\langle F_x(t_1)\,F_x(t_2) \rangle = C\, \delta (t_1-t_2),
\end{equation}
where $\delta$ is a Dirac delta function and the angular brackets denote an
average over an ensemble of ``similarly prepared'' systems of stars in each
of which the black hole has the same initial position and velocity. We take
the factor $C$ to be independent of $r$; its magnitude will be determined
in the next section. While this definition will not allow us to solve
equation (\ref{eqofmotion}) explicitly, we will obtain closed expressions
for the time autocorrelation function of the black hole position and
velocity in the next section.
That the components of the random force can be separated and characterized
as in the latter part of equation (\ref{stochdef}) is at this stage an
assumption; its justification must ultimately come from the agreement
between the results of the model and the numerical simulations, as detailed
in \S 5.

The autocorrelation function of the stochastic force on an individual star
has been calculated before (Chandrasekhar 1944a,b; Cohen 1975), in the
approximation that the test star and its surrounding stars move along
straight lines on deterministic orbits; in this approximation, the
autocorrelation function falls off as slowly as the inverse of the time lag
for a uniformly dense infinite system. However, for the case we study in
this paper, the fall-off will be much faster as a consequence of the rapid
decrease in the density of the system outside the core radius (see Cohen
1975), and because fluctuations will tend to throw the black hole and the
field stars off their deterministic paths and by doing so reduce the
correlation (see Maoz 1993). Another difference arises from the fact that
we consider here a test object which is much more massive than the
surrounding stars; since the black hole moves very slowly relative to the
stars, in the time that the motion of the black hole changes appreciably,
the correlations in the force due to the stars would have worn off. Our
choice of the delta function to represent the force autocorrelation
function is somewhat of an idealization, but is justified \emph{a
posteriori} by the good agreement between the model outlined above and the
results of simulations described in \S 5.

Before going on to solve the equations of motion, it is useful to list the
approximations that have gone into setting up our model. We have assumed
that a black hole of mass $m$ is located near the center of a stellar
system of total mass $M$ and characteristic length scale $a$, and that the
mass of individual stars $m_{\star} \ll m$.  Hence, the black hole's
velocity is expected to be very small compared with the velocities of the
stars, and its position is expected to be confined in a small region, $r
\ll a$.  We assume that the total force on the black hole is made up of two
independent, separable parts.  One (i.e., ${\bf F}(t)$), which is due to
very rapid fluctuations in the immediate surroundings of the black hole, is
assumed to average to zero and to be uncorrelated with itself.  The other
part, which consists of dynamical friction and the force due to the
aggregate stellar system, varies smoothly with the black hole's position
and velocity on a time scale very much longer than that of the
fluctuations.

We assume that dynamical friction is given by the Chandrasekhar
formula, which entails a number of additional approximations (see
Tremaine \& Weinberg 1984; Weinberg 1986; Nelson \& Tremaine 1999).
The Chandrasekhar formula was originally derived for an infinite and
homogeneous stellar system, but it is often employed for
non-homogeneous systems by replacing the homogeneous density by the
local density. The maximum effective impact parameter (for relaxation
encounters between the stars and the black hole) that enters the
Coulomb logarithm is not well-defined; we assume it to be given
implicitly by the density-weighted expression (\ref{bmax}) above. The
gravitational encounters between the stars and the black hole are
treated as a succession of binary encounters of short duration,
i.e. as a Markov process. The Chandrasekhar formula approximates the
orbits on which stars move past the black hole as Keplerian
hyperbolae, even though the actual stellar orbits are more
complex. This formula neglects the self-gravity of stars in the wake
induced by the black hole.  Despite these approximations,
Chandrasekhar's formula has been found to provide an accurate
description of dynamical friction in a variety of astrophysical
situations (see BT and references therein).  In the present context,
we will gauge the reliability of its use by appealing to numerical
simulations to test the applicability of our model.

We conclude this section by demonstrating that the time-scale for
fluctuations in $F_x(t)$ is very much smaller than the time-scale on
which the position and velocity of the black hole change.

Near the center of a Plummer model, where the massive black hole is
localized, the stellar density is $\rho \sim ({3M}/{4 \pi
a^3})$, since $r \ll a$; therefore, the typical separation between
stars is $D \sim \left({4 \pi a^3 m_{\star}}/{3M}
\right)^{1/3}$. The typical stellar velocity is $V_0 \sim \left( 
GM/2a \right)^{1/2}$. The average time period of
changes in $F_x(t)$, caused by discrete stellar encounters,
is then approximately $T_{loc} \sim
D/V_0$. Now the characteristic time period with which the black
hole's motion changes is $T_{BH} \sim 2 \pi / \omega_0$,
where $\omega_0 = \sqrt{GM/a^3}$, as is shown in the next section. We
thus have
\begin{displaymath}
\frac{T_{loc}}{T_{BH}} \sim \left( \frac{\sqrt{2} m_{\star}}
{3 \pi^2 M} \right)^{1/3} = \left( \frac{\sqrt{2}}{3 \pi^2 N} 
\right)^{1/3} = \frac{0.36}{N^{1/3}},
\end{displaymath}
if there are a total of $N$ stars each of mass $m_{\star}$ (i.e.,
$M=m_{\star} N$). Therefore, $T_{loc} \ll T_{BH}$ for large $N$, and 
we are justified in separating the total force on the black hole into 
slowly varying and rapidly fluctuating contributions.

\section{Solution of the Model Equations}

If we choose the initial position and velocity of the black hole to be 
\begin{equation} \label{initcond}
x(0)=x_0, \qquad \dot{x}(0)=v_0,
\end{equation}
then we can solve equation (\ref{eqofmotion}) formally as 
\begin{equation} \label{x}
x(t)=e^{-\gamma t} \Big( x_0 \cos \omega t + \frac{v_0+\gamma x_0}
{\omega} \sin \omega t \Big) + \frac{1}{m \omega} \int_0^t F_x (t-z)\,
e^{-\gamma z} \sin \omega z\, dz.
\end{equation}
Using Leibniz's rule to differentiate the second term on the right
hand side of equation (\ref{x}) under the integral sign, we can also
solve for the velocity:
\begin{equation} \label{v}
\dot{x}(t)=e^{-\gamma t} \Big( v_0 \cos \omega t - \frac{\gamma v_0+ 
{\omega_0}^2 x_0}
{\omega} \sin \omega t \Big) + \frac{1}{m \omega} \int_0^t F_x (t-z)\,
e^{-\gamma z} (- \gamma \sin \omega z + \omega \cos \omega z)\, dz.
\end{equation}
In the above equations, 
\begin{equation} \label{defs}
\omega_0=\sqrt{\frac{k}{m}}, \qquad \gamma = \frac{\beta}{2m}, \qquad
\omega = \sqrt{\omega_0^2 - \gamma^2}.
\end{equation}
In the case of interest with $m \ll M$, we have $\gamma \ll \omega_0$,
and so $\omega \simeq \omega_0$. 
Note that the exact results of equation (\ref{defs}) -- and not the
above approximations -- have been used in comparing the predictions of
the model with the numerical simulations of \S 5.

Using the first of the properties in equation (\ref{stochdef}), we
have the following:
\begin{equation} \label{xav}
\langle x(t) \rangle =x_0 e^{-\gamma t} \Big( \cos \omega t + 
\frac{\gamma}{\omega} 
\sin \omega t \Big) + \frac{v_0}{\omega} e^{-\gamma t} \sin \omega t,
\end{equation}
\begin{equation} \label{vav}
\langle \dot{x}(t) \rangle =v_0 e^{-\gamma t} \Big( \cos \omega t - 
\frac{\gamma}{\omega} 
\sin \omega t \Big) - x_0 \frac{\omega_0^2}{\omega} e^{-\gamma t} 
\sin \omega t.
\end{equation}
Note that in the steady state (i.e., as $t \rightarrow \infty$), the
average values of the position and velocity components are zero. 
In the above equations and the subsequent equations, angular brackets have
the same meaning as in equation (\ref{stochdef}).

Using the second of the properties in equation (\ref{stochdef}), we
can employ the delta function to perform the resulting double integral
and solve for the time autocorrelation functions of the black hole
position and velocity, with a time lag $T$:
\begin{eqnarray} \label{xcorrfull}
\langle x(t) x(t+T) \rangle  & = & \langle x(t) \rangle \langle 
x(t+T) \rangle {} \nonumber\\  
& & {} + \frac{C e^{-\gamma T}}{4 m^2 \omega^2 \omega_0^2} \lbrack 
\omega \sin \omega T \lbrace 1-e^{-2 \gamma t} (1-2 \sin^2 \omega t
+ \frac{\gamma}{\omega} \sin 2 \omega t) \rbrace {} \nonumber\\
& & {}+ \frac{\omega^2}{\gamma} \cos \omega T \lbrace 1-e^{-2 
\gamma t} (1+2 \frac{\gamma^2}{\omega^2}\sin^2 \omega t
+ \frac{\gamma}{\omega} \sin 2 \omega t) \rbrace \rbrack,
\end{eqnarray}
\begin{eqnarray} \label{vcorrfull}
\langle \dot{x}(t) \dot{x}(t+T) \rangle  & = & \langle \dot{x}(t) 
\rangle \langle \dot{x}(t+T) \rangle {} \nonumber\\  
& & {} + \frac{C e^{-\gamma T}}{4 \gamma m^2} \lbrack \frac{\gamma}
{\omega}\sin \omega T\lbrace -1+e^{-2 \gamma t} (1-2 \sin^2 \omega t
- \frac{\gamma}{\omega} \sin 2 \omega t) \rbrace {} \nonumber\\
& & {}+ \cos \omega T \lbrace 1-e^{-2 
\gamma t} (1+2 \frac{\gamma^2}{\omega^2}\sin^2 \omega t
- \frac{\gamma}{\omega} \sin 2 \omega t) \rbrace \rbrack.
\end{eqnarray}
In the same way we can calculate another quantity that will be useful
later:
\begin{equation} \label{xvav}
\langle x(t) \dot{x}(t) \rangle = \langle x(t) \rangle \langle \dot{x}
(t) \rangle + \frac{C}{2 m^2 \omega^2} e^{-2 \gamma t} \sin^2 \omega t.
\end{equation}
Using equations (\ref{xcorrfull}) and (\ref{vcorrfull}), we can obtain the
corresponding steady state autocorrelation functions for position and
velocity, which are functions of the time lag $T$ alone, by letting $t$ go
to infinity and thus letting transients die out,
\begin{equation} \label{xcorr}
C_{xx}(T) \equiv \lim_{t \rightarrow \infty} \langle x(t) x(t+T) 
\rangle = \frac{C e^{-\gamma T}}{4 \gamma m^2 \omega_0^2} \lbrack
\cos \omega T + \frac{\gamma}{\omega} \sin \omega T \rbrack,
\end{equation}
\begin{equation} \label{vcorr}
C_{v_x v_x}(T) \equiv \lim_{t \rightarrow \infty} \langle \dot{x}(t)
\dot{x}(t+T) 
\rangle = \frac{C e^{-\gamma T}}{4 \gamma m^2} \lbrack
\cos \omega T - \frac{\gamma}{\omega} \sin \omega T \rbrack.
\end{equation}
Since, in most cases of interest, $\gamma \ll \omega$, these functions
are essentially pure damped cosine terms with zero phase.
Note that the stationary state autocorrelation functions are related
by
\begin{equation} \label{corrrel}
C_{v_x v_x}(T) = - \frac{d^2 C_{xx}(T)}{d T^2}.
\end{equation}

It remains now to determine the constant $C$. If we multiply
(\ref{eqofmotion}) by $\dot{x}$, rearrange and take the ensemble
average, we obtain
\begin{displaymath}
\langle \frac{d}{dt}(\frac{1}{2}m \dot{x}^2+\frac{1}{2}k x^2) \rangle
= \langle F_x(t) \dot{x} \rangle - \langle \beta \dot{x}^2 \rangle .
\end{displaymath}
In the steady state, the average rate of change of total energy of the
black hole is zero, and we get
\begin{equation} \label{fluctdiss}
\langle F_x(t) \dot{x} \rangle = \langle \beta \dot{x}^2 \rangle ;
\end{equation}
in other words, the ``heating'' by the medium due to fluctuations must
in the steady state equal the ``cooling'' due to viscous dissipation
by the force of dynamical friction, which is a form of the general
relationship between the processes of fluctuation and dissipation (see
Bekenstein \& Maoz 1992; Maoz 1993; Nelson \& Tremaine 1999). 

The value of $\langle \dot{x}^2 \rangle$ in the steady state is easily 
evaluated by setting $T=0$ in equation (\ref{vcorr}); thus,
\begin{equation} \label{cooling}
\langle \beta \dot{x}^2 \rangle = \frac{C \beta}{4 \gamma m^2}.
\end{equation}

BT calculate the total heating per unit time to be $16 {\pi}^2 {\rm ln}
\Lambda\, G^2 m m_{\star} \int_{v}^{\infty} u f(r,u) du$
[adapted from equation (8-66) in BT].  
Isotropy implies that the heating
due to the $x$-component alone will be a third of this quantity, namely
\begin{displaymath}
\langle F_x(t) \dot{x} \rangle = \frac{16}{3} {\pi}^2 {\rm
ln} \Lambda\, G^2 m m_{\star} \int_{v}^{\infty} u f(r,u) du .
\end{displaymath}
Since the black hole velocity $v$ is small, we can replace the lower limit
in the integral above by zero. Then, for the Plummer model we obtain
\begin{displaymath}
\int_{v}^{\infty} u f(r,u) du \simeq
C_1 \int_{0}^{\sqrt{2 \Psi}} u \bigg( \Psi (r) - \frac{u^2}{2}
\bigg)^{7/2}\, du = \frac{2}{9} C_1 \Psi (r)^{9/2} \simeq \frac{2}{9}
C_1 \bigg( \frac{GM}{a} \bigg)^{9/2} \quad \textrm{for small $r$},
\end{displaymath}
where 
\begin{displaymath}
C_1 = \frac{96}{7 \sqrt{8} {\pi}^3} \frac{M a^2}{(G M)^5} .
\end{displaymath}
By plugging this back into the expression for $\langle F_x(t)
\dot{x} \rangle$ and using equations (\ref{cooling}), 
(\ref{fluctdiss}) and (\ref{b}), we obtain finally
\begin{equation} \label{C}
C = \frac{8}{9} \frac{GM}{a} \gamma m m_{\star},
\end{equation}
and
\begin{displaymath}
\langle \dot{x}^2 \rangle = \frac{C}{4 \gamma m^2} = \frac{2}{9} 
\frac{GM}{a}\frac{m_{\star}}{m}.
\end{displaymath}
The first equality in the above equation was obtained from equation
(\ref{vcorr}) by setting $T=0$. Note that this is slightly higher than the
value that would have been obtained had the black hole's kinetic energy
been in strict equipartition with that of the stars in the core of the
Plummer potential; had that been the case, the numerical coefficient above
would have been 1/6 instead of 2/9 (see equation (\ref{msvstar}), where the
mean squared 3-dimensional velocity of the stars in the core has been
calculated).

As an aside, we point out that a similar calculation for a Maxwellian
distribution of stars --- $f \propto e^{-v^2/2 \sigma^2}$, where 
$\sigma$ is the root mean squared value of a single component of 
velocity --- would have yielded 
\begin{displaymath}
C = 4 \gamma m m_{\star} \sigma^2,
\end{displaymath}
and 
\begin{displaymath}
\langle \dot{x}^2 \rangle = \frac{C}{4 \gamma m^2} = \frac{m_{\star} 
\sigma^2} {m}.
\end{displaymath}
This is the familiar condition for equipartition between the kinetic
energies of the black hole and a star.

Returning to the Plummer model, we have, by making use of equations
(\ref{C}), (\ref{xcorr}) and (\ref{vcorr}), the following expressions for
the mean squared position and velocity components of the black hole in
the steady state:
\begin{equation} \label{rmsx}
\langle x^2 \rangle = \frac{2}{9} \frac{a^2 m_{\star}}{m},
\label{eq:x^2}
\end{equation}
\begin{equation} \label{rmsv}
\langle v_x^2 \rangle = \frac{2}{9} \frac{GM}{a} \frac{m_{\star}}{m};
\label{eq:v_x^2}
\end{equation}
(compare equation (\ref{rmsx}) to equation (101) in Bahcall \& Wolf 1976; 
the latter is rederived in Lin \& Tremaine 1980 just after their
equation (16); these equations agree with our result above to within a 
factor of order unity).
If we have $N$ stars in the cluster of total mass $M$, then $m_{\star} 
=M/N$, and we can rewrite the above equations as
\begin{equation} \label{rmsxN}
\langle x^2 \rangle = \frac{2}{9} \frac{a^2 M}{m N},
\end{equation}
\begin{equation} \label{rmsvN}
\langle v_x^2 \rangle = \frac{2}{9} \frac{GM^2}{a m N} .
\end{equation}

\section{The Probability Distributions of the Black Hole Position and
Velocity}

Following the treatment of Chandrasekhar (1943a) and Wang and
Uhlenbeck (1945), we can derive a partial differential equation,
called the Fokker-Planck equation, for the joint probability
distribution of the position and velocity components of the black
hole. 

Let $W(x,v_x,t)$ represent the probability distribution of the $x$-
components of the black hole's position and velocity at time $t$; i.e.,
$W(x,v_x,t)\, \Delta x\, \Delta v_x$ is the probability that at time $t$,
the black hole lies between $x$ and $x+\Delta x$ and has a velocity between
$v_x$ and $v_x+\Delta v_x$. Let $\psi(x+\Delta x, v_x+\Delta v_x,t+\Delta t
|x,v_x,t)$ represent the (conditional) transition probability that at time
$t+ \Delta t$, the black hole is at $x+\Delta x$ and $v_x+\Delta v_x$,
given that at time $t$, it was at $x$ and $v$; $\Delta t$ is taken to be an
interval that is long compared with the time-scale over which the
stochastic force $F_x(t)$ varies but is short compared with the time-scale
on which the black hole's position and velocity change.

The evolution of the probability $W$ is expected to be governed by the
following equation:
\begin{displaymath} 
W(x,v_x,t+\Delta t) = \int W(x-\Delta x,v_x-\Delta v_x,t)
\psi(x,v_x,t+\Delta t |x-\Delta x,v_x-\Delta v_x,t) d(\Delta x)
d(\Delta v_x)
\end{displaymath}
Note that in writing this equation, we are assuming that the black
hole's motion is a Markov process which depends only on its position
and velocity an ``instant'' before, and is independent of its previous
history. Rewriting the expression for $\psi$ in the above equation as
$\psi(x-\Delta x+\Delta x,v_x-\Delta v_x+\Delta v_x,t+\Delta t 
|x-\Delta x,v_x-\Delta v_x,t)$, and expanding both sides of the
equation in Taylor series, we have
\begin{displaymath}
W(x,v_x,t)+\frac{\partial W}{\partial t} \Delta t + \cdots  = 
\int \sum_{i=0}^{\infty} \frac{(-1)^i}{i!}
\Delta y_1 \ldots \Delta y_i \frac{\partial^i}{\partial y_1 \ldots
\partial y_i} \lbrack \psi_{\Delta} 
W(x,v_x,t) \rbrack d(\Delta x) d(\Delta v_x), 
\end{displaymath}
where each $y_i$ is either $x$ or $v_x$, and where for brevity we 
have defined 
\begin{displaymath}
\psi_{\Delta} \equiv \psi(x+\Delta x,v_x+\Delta v_x,t+\Delta t 
|x,v_x,t).
\end{displaymath}
Keeping only derivatives up to the second order on the right hand
side, we have
\begin{eqnarray*}
W(x,v_x,t)+\frac{\partial W}{\partial t} \Delta t + \cdots & = &
\int d(\Delta x) d(\Delta v_x) \lbrack 1 - \Delta x \frac{\partial}
{\partial x} - \Delta v_x \frac{\partial}{\partial 
v_x} \\
& &
+\frac{1}{2}(\Delta x)^2 
\frac{\partial^2}{\partial x^2} +\frac{1}{2}(\Delta v_x)^2 
\frac{\partial^2}{\partial v_x^2}+\Delta x\, \Delta v_x 
\frac{\partial^2}{\partial x\, \partial v_x} \rbrack
(\psi_{\Delta} W).
\end{eqnarray*}
The first term on the right hand side is simply $W(x,v_x,t)$, which
cancels with the same term on the left hand side. Dividing both sides
by $\Delta t$ and taking the limit $\Delta t \rightarrow 0$, we obtain
\begin{equation} \label{FP}
\frac{\partial W}{\partial t} = -\frac{\partial}{\partial x}(D_x W)
-\frac{\partial}{\partial v_x}(D_{v_{x}} W)
+\frac{1}{2}\frac{\partial^2}{\partial x^2}(D_{xx} W)
+\frac{1}{2}\frac{\partial^2}{\partial v_x^2}(D_{v_x v_x} W)
+\frac{\partial^2}{\partial x\, \partial v_x}(D_{x v_x} W),
\end{equation}
where the $D$ coefficients are the diffusion coefficients of this
general Fokker-Planck equation in two variables, and are defined as
\begin{displaymath}
D_x \equiv \lim_{\Delta t \rightarrow 0} \frac{1}{\Delta t}
\int\, d(\Delta x) d(\Delta v_x) (\Delta x) \psi_{\Delta}
=\lim_{\Delta t \rightarrow 0} \frac{\langle \Delta x \rangle}
{\Delta t},
\end{displaymath}
\begin{displaymath}
D_{xx} \equiv \lim_{\Delta t \rightarrow 0} \frac{1}{\Delta t}
\int\, d(\Delta x) d(\Delta v_x) (\Delta x)^2 \psi_{\Delta}
=\lim_{\Delta t \rightarrow 0} \frac{\langle (\Delta x)^2 \rangle}
{\Delta t},
\end{displaymath}
\begin{displaymath}
D_{x v_x} \equiv \lim_{\Delta t \rightarrow 0} \frac{1}{\Delta t}
\int\, d(\Delta x) d(\Delta v_x) (\Delta x \Delta v_x) 
\psi_{\Delta}
=\lim_{\Delta t \rightarrow 0} \frac{\langle \Delta x \Delta v_x 
\rangle}{\Delta t}, \quad \textrm{etc.}
\end{displaymath}

The diffusion coefficients can be calculated very easily by using the
equation of motion (\ref{eqofmotion}) and the definition of the
autocorrelation of the random force $F_x(t)$ in equations
(\ref{stochdef}). We have
\begin{displaymath}
\Delta x = v_x \Delta t,
\end{displaymath}
and by integrating the equation of motion for a short time $\Delta t$
which is long enough that many random encounters have taken place but
not so long that the black hole's $x$ and $v_x$ have changed
appreciably,
\begin{displaymath}
\Delta v_x = - \left( \frac{b}{m} v_x + \frac{k}{m} x \right) 
\Delta t + \int_t^{t+\Delta t} \frac{F_x(t)}{m}\, dt.
\end{displaymath}
Based on the above and equations (\ref{stochdef}), we find that the only
diffusion coefficients that do not vanish as $\Delta t \rightarrow 0$
are:
\begin{displaymath}
D_x=v_x, \quad D_{v_x}=-\left( \frac{b}{m} v_x + \frac{k}{m} x 
\right), \quad \textrm{and} \quad D_{v_x v_x}=\frac{C}{m^2}.
\end{displaymath} 
Thus, the Fokker-Planck equation reduces to
\begin{equation} \label{FPred}
\frac{\partial W}{\partial t} = -\frac{\partial}{\partial x}(v_x W)
+ \frac{\partial}{\partial v_x} \Biggl\lbrack \left( \frac{b}{m} v_x + 
\frac{k}{m} x \right) W \Biggr\rbrack
+\frac{C}{2 m^2} \frac{\partial^2 W}{\partial v_x^2}.
\end{equation}

The stationary distribution $W_{st}(x,v_x) \equiv W(x,v_x,t \rightarrow
\infty)$ is found by setting the time derivative on the left hand side of
equation (\ref{FPred}) to zero.

The solution of equation (\ref{FPred}) is complicated, but we write it
down in terms of the quantities derived in previous sections (see
Chandrasekhar 1943a):
\begin{equation} \label{FPsol}
W(x,\dot{x},t)=\frac{1}{2 \pi \sqrt{D_W}} \textrm{exp} \lbrack -
\lbrace A_W \left( x-\langle x \rangle \right)^2 +
B_W \left( \dot{x}-\langle \dot{x} \rangle \right)^2 -
2 C_W \left( x-\langle x \rangle \right) \left( \dot{x}-\langle
\dot{x} \rangle \right) \rbrace \rbrack,
\end{equation}
where 
\begin{displaymath}
D_W = \left( \langle x^2 \rangle - \langle x \rangle^2 \right)
\left( \langle \dot{x}^2 \rangle - \langle \dot{x} \rangle^2 \right)
-\Big( \langle x \dot{x} \rangle -
\langle x \rangle \langle \dot{x} \rangle \Big)^2,
\end{displaymath}
\begin{displaymath}
A_W = \frac{\langle \dot{x}^2 \rangle - \langle \dot{x} \rangle^2}
{2 D_W},
\end{displaymath}
\begin{displaymath}
B_W = \frac{\langle x^2 \rangle - \langle x \rangle^2}{2 D_W},
\end{displaymath}
\begin{displaymath}
C_W = \frac{\langle x \dot{x} \rangle - \langle x \rangle \langle 
\dot{x} \rangle}{2 D_W}.
\end{displaymath}
Note that equation (\ref{FPsol}) describes a general Gaussian
distribution in the two variables $x$ and $\dot{x} \equiv v_x$.

Of particular interest is the stationary distribution $W_{st}
(x,v_x)$, which is obtained from equation (\ref{FPsol}) by taking the 
limit $t \rightarrow \infty$:
\begin{equation} \label{FPsolst}
W_{st}(x,v_x)= \frac{2 \gamma m^2 \omega_0}{\pi C}
\textrm{exp} \Biggl\lbrack - \frac{2 \gamma m^2}{C} \left( \omega_0^2 
x^2 + v_x^2 \right) \Biggr\rbrack;
\end{equation}
this is the product of two \emph{independent} Gaussian distributions 
in the variables $x$ and $v_x$. 
This is a consequence of the linear nature of equations
(\ref{veceqofmotion}) and (\ref{eqofmotion}).
It is easy to obtain the marginal
stationary distributions of these variables by integrating out one or
the other:
\begin{equation} \label{FPsolstx}
W_{st}(x)={1\over \sqrt{2\pi \langle x^2\rangle}} \,\, 
\textrm{exp} \Biggl\lbrack - \frac{x^2}{2\langle x^2\rangle} 
\Biggr\rbrack,
\end{equation}
\begin{equation} \label{FPsolstv}
W_{st}(v_x)={1\over \sqrt{2\pi \langle v_x^2\rangle}} \,\, 
\textrm{exp} \Biggl\lbrack - \frac{v_x^2}{2\langle v_x^2\rangle} 
\Biggr\rbrack,
\end{equation}
where $\langle x^2\rangle$ and $\langle v_x^2\rangle$ are given by 
equations~(\ref{eq:x^2}) and (\ref{eq:v_x^2}), respectively.
%

\section{Tests of the Model using Numerical Simulations}

We have performed a number of computer simulations to test the validity of
the model presented in \S 2--4.  The code we use solves the combined
dynamics of the black hole and the stars using different equations of
motion:

\begin{equation}                                    \label{eq-bh} 
m \ddot{{\bf r}} =
\sum_k { G m m_{\star,k} ({\bf r}_{\star,k}-{\bf r}) \over
             ( |{\bf r}_{\star,k}-{\bf r}|^2 + \epsilon^2 )^{3/2} },
\end{equation}
\begin{equation}                                    \label{eq-st}
m_{\star,k} \ddot{{\bf r}}_{\star,k} = {G m m_{\star,k} ({\bf r}-
{\bf r}_{\star,k}) \over
( |{\bf r}-{\bf r}_{\star,k}|^2 + \epsilon^2 )^{3/2}} 
- m_{\star,k} \nabla \Phi (r_{\star,k});
\end{equation}
where $m$ and ${\bf r}$ are the mass and position of the black hole,
respectively, and $m_{\star,k}$ and ${\bf r}_{\star,k}$ are the mass and
position of the $k$-th star, respectively; the Coulomb force is softened by
the parameter $\epsilon$ to prevent numerical divergences when a star
passes very close to a black hole; and $\Phi (r)$ is the analytical
expression for the stellar potential in equation (\ref{pot}). Thus, the
black hole interacts with the stars through a softened Coulomb force, and
the stars interact with each other through an analytical gravitational
field.  Combining an analytical potential with the traditional ``direct
summation'' N-body technique ensures that accuracy is not sacrificed in
calculating the motion of the black hole (the object of greatest interest
for us).  The particles themselves are moved (with varying step-sizes which
are calculated at every time step) using suitably modified versions of the
fourth-order integrators of Aarseth (1994).

The improved efficiency in the calculation is thus obtained at the price of
having to keep the potential due to the stars (although not necessarily
their density profile) fixed. However, this approximation does not appear
to have a significant effect on our results. We have performed a number of
simulations using other methods to test the results. These include the
direct summation N-body code known as NBODY1 (Aarseth 1994) for a
relatively small number of particles, and the program known as SCFBDY,
described in detail in Quinlan \& Hernquist (1997). The latter program
expresses the potential as an expansion in an appropriate set of basis
functions instead of having a fixed potential in equation (\ref{eq-st})
above; the coefficients of this expansion are self-consistently updated at
chosen time steps. Although the precise motion of the black hole is not
identical for different simulation methods -- since the force on the black
hole in each case is calculated differently -- we have found that they all
give similar results as far as the statistical properties of the black
hole's dynamics are concerned. In particular, the mean squared values of
the black hole's position and velocity in the stationary state of the
system are approximately equal irrespective of the method used, and are
similar to the values derived from the model presented in this paper. We
believe that this is because the statistical properties of the black hole's
motion are determined primarily by the properties of the restoring force
and dynamical friction which are provided by the unbound stars, outside the
region of the black hole's gravitational influence. These regions are
relatively unaffected by the central black hole if its mass is much smaller
than the total mass of the stellar system. That being so, we have used the
method of the fixed potential for the simulations described below in order
to be able to integrate efficiently large numbers of stars for long spans
of time.

Our standard Plummer model has parameters $G=M=1$ and $a=3 \pi /16$;
in these units, the gravitational energy of the initial stellar
system alone is $-1/4$, and the circular period at $r=a$ is
$t_{circ}\approx 4.78$. We take the mass of the black hole to be 
$m=0.01$. The softening length was chosen to be
$\epsilon = 5 \times 10^{-3}$. 
For these parameters, $\omega_0=2.212, \gamma=0.30$ and 
$\omega=2.192$.

In Figure 1, we show the results of a simulation in which the black hole
was started off with zero velocity from the origin in a system of
$N=100,000$ stars.  The first and second panels show the evolution of the
black hole's x-component of position and velocity, respectively. The third
panel shows the autocorrelation function of the x-component of the black
hole's position as calculated from the simulation (the calculation was
stopped at time $t=600$), and as computed from our model; the two curves
are in good agreement, at least for short time lags, and the discrepancies
could be due to the uncertainty in the maximum effective impact parameter
in the dynamical friction formula.  Note the persistence of the actual
autocorrelation function of the black hole, which will be discussed further
below.  The autocorrelation function of $v_x$ is not shown; according to
equation (\ref{corrrel}) it can be simply derived from the autocorrelation
function of $x$ by taking a double time derivative.

In Figure 2 we test equations (\ref{rmsxN}) and (\ref{rmsvN}) which
predict that the root mean squared position and velocity components of
the black hole should decline with the total number of stars $N$ as
$N^{-1/2}$.  We show the results of 4 simulations with $N=\, $12,500,
25,000, 50,000 and 100,000; in each case, the simulation was stopped
at time $t=600$; the agreement with the predictions of the
model is evidently good. 

In Figure 3, we test equations (\ref{FPsolstx}) and (\ref{FPsolstv}), which
predict that the black hole's position and velocity components in the
steady state should be Gaussian distributed; the empirically binned
distributions were computed for the case with $N=100,000$. The agreement
with the model predictions is again very good.

In the above simulations, the black hole's orbit remains close to the
center and appears to be essentially stochastic, in that it does not seem
to be confined to a special sheet or line in phase space. At any point in
time, many stars are bound to the black hole in the sense that they have
negative energy with respect to it. Most of these stars are within the
gravitational sphere of influence of the black hole and their total mass is
comparable to that of the black hole.

The autocorrelation functions of the black hole's position and velocity do
not appear to damp entirely with ever increasing time lag $T$, in
contrast to equations (\ref{xcorr}) and (\ref{vcorr}).  Although Figure 1
shows only a small part of the autocorrelation function, it turns out that
the oscillations persist for indefinitely long $T$ at roughly the residual
(and, apparently, somewhat varying) amplitude shown at the right of the
third panel in the figure. The frequency of the oscillations is close to
the fundamental frequency calculated in the paper and evident in the third
panel of Figure 1.  We attribute these oscillations to the presence of very
weakly damped coherent modes in the stellar system, of the kind reported by
Miller (1992) and Miller \& Smith (1992), and calculated by Mathur (1990)
and Weinberg (1994).

While it is not the purpose of this paper to study such modes, we have, in
an attempt to identify the source of the above oscillations, performed the
following experiment. We set up the system of stars as above but
\emph{without} the black hole at its center, and kept track of one
component of the total force at the origin of the system. The discrete
Fourier transform of the sequence of forces at successive time steps
revealed a strong peak very close to the fundamental angular frequency of
oscillations at the bottom of the gravitational potential well: $\omega_0
\equiv \sqrt{k/m} = \sqrt{GM/a^3}$.  This could account for the undamped
low-amplitude oscillations, at roughly the above frequency, seen in the
autocorrelation functions of the black hole's position and
velocity. Consider a simplified situation in which the black hole is
subject to an additional force $D_1 \cos (\omega_0 t) + D_2 \sin (\omega_0
t)$, which is due to the conjectured undamped mode of this frequency
mentioned above; here, $D_1$ and $D_2$ are taken to be independent of time,
for simplicity. If we add this force to the right hand side of equation
(\ref{eqofmotion}), assume that it is independent of the random force
$F_{x}(t)$, and perform an analysis similar to that in \S 3, it is
easy to see that we would obtain a new contribution to each of the
autocorrelation functions in equations (\ref{xcorr}) and (\ref{vcorr}) in
the form of an additive term which is proportional to $\cos (\omega_0 T)$,
i.e., a term that does not damp with increasing $T$ (note that for most
systems, $\omega_0$ is approximately equal to $\omega = \sqrt{\omega_0^2 -
\gamma^2}$, the frequency with which the black hole's position and velocity
autocorrelation functions oscillate).  The addition of such a term would
not affect the good agreement for small time lags between equations
(\ref{xcorr}) and (\ref{vcorr}) and the results of numerical simulation
(Figure 1), since the amplitude of these residual oscillations is very much
smaller than the amplitude of the autocorrelation functions for small $T$.

\section{Lower Limit on the Mass of Sgr A*}

We may use our model to derive a lower limit on the mass of the black hole
in the Galactic center, Sgr A*. The observed upper limit of 20 km s$^{-1}$
on the intrinsic proper motion of Sgr A* (Reid et al. 1999), when combined
with equation (\ref{rmsv}), provides such a limit.

Measurement of proper motions of stars close to Sgr A* indicate that a
total mass of $\sim 2.6 \times 10^6 M_{\odot}$ resides within a distance of
$\sim 0.01$ parsec of Sgr A* (see, e.g., Eckart \& Genzel 1997, Ghez et
al. 1998, Ghez et al. 2000). Not all of this mass need be attributed to the
black hole; some of it could be due to a cluster of stars surrounding
it. If we assume that this cluster is distributed according to a Plummer
profile, then we can apply the results derived in previous sections to the
entire system comprised of the stellar cluster and the black hole.

Let us set the total mass inside a distance $R = 0.01$ pc from 
Sgr A* to be $M_R=2.6 \times 10^6 M_{\odot}$. Using $m$ as
the mass of the black hole, we have the condition 
\begin{equation} \label{MR}
M_R = m + \frac{M R^3}{(R^2+a^2)^{3/2}},
\end{equation}
where the second term is the mass of the stellar cluster (of total mass
$M$) within $R$. Combining this with equation (\ref{rmsv}) in the form
\begin{displaymath}
\langle v_x^2 \rangle = \frac{2}{9} \frac{GM}{a} \frac{m_{\star}}{m} <
\langle v_x^2 \rangle_{max}, 
\end{displaymath}
we obtain
\begin{equation} \label{mlolim}
m > {M_R\over 1+\frac{9}{2} \frac{\langle v_x^2\rangle_{max} a R^3}{G 
(R^2+a^2)^{3/2} m_{\star}}},
\end{equation}
where $<v_x^2>_{max}$ is the maximum mean squared speed of one component of
the black hole's velocity. We can now set $\langle v_x^2
\rangle_{max}^{1/2} \sim 20$ km s$^{-1}$ to get an approximate lower limit
on $m$, assuming that $m_{\star} \sim 1 M_{\odot}$. This relation is
plotted in Figure 4(a) for various values of $a$, the scale length
parameter of the Plummer cluster. The mass of the black hole must be given
by points lying \emph{above} the curved solid line.  Evidently, this
relation implies a lower limit on $m$.

We can derive stricter limits on $m$ by noting, as did Maoz (1998),
that the allowed values of $a$ are restricted by the condition that 
the upper limit of the lifetime of a cluster of stars is its
evaporation time, when stars would have escaped from the cluster 
because of scattering. The evaporation timescale of the cluster should
be long enough to make it probable that the cluster be observed at the
present epoch. A reasonable assumption is that this timescale,
$t_{evap}$, is bounded between the values 1 Gyr and 10 Gyr, the latter
being the approximate age of the Galaxy. The evaporation timescale is
$t_{evap} \approx 300 t_{rh}$, where $t_{rh}$ is the median
relaxation time, given by (see, e.g., Maoz 1998, BT)
\begin{displaymath}
t_{rh}=\frac{0.14 N}{\ln(0.4 N)} \sqrt{\frac{r_h^3}{GM}},
\end{displaymath}
where $N=M/m_{\star}$ is the number of stars in the cluster (with $M$ given
by equation~\ref{MR}), and $r_h$ is the system's median radius; $r_h=1.3a$
for the Plummer profile.

Setting $t_{evap}$ = 1 Gyr, we obtain the dashed line in Figure 4(a);
$t_{evap} >$ 1 Gyr denotes the region to the \emph{right} of this
line. Hence, the allowed values of $m$ and $a$ lie to the right of this
dashed line and above the solid line. The minimum value of $m$ under these
assumptions is given by point A at which $m \approx 3.8 \times 10^3
M_{\odot}, a \approx 0.021$ pc, and the total mass of the star cluster is
$M \approx 3.3 \times 10^7 M_{\odot}$.

If we perform a similar calculation for $t_{evap} >$ 10 Gyr, we obtain the
dotted line in Figure 4(a); the minimum value of $m$ is then given by point
B at which $m \approx 1.6 \times 10^4 M_{\odot}, a \approx 0.049$ pc and $M
\approx 3.3 \times 10^8 M_{\odot}$.

Reid et al. (1999) expect further observations to reduce the limit on the
peculiar motion of Sgr A* from 20 km s$^{-1}$ to 2 km s$^{-1}$ over the
next few years. Figure 4(b) repeats the above calculations for this
limit. The condition $t_{evap} >$ 1 Gyr then gives a new lower limit for
$m$ (point A), at which $m \approx 3.4 \times 10^5 M_{\odot}, a \approx
0.022$ pc, and $M \approx 3.0 \times 10^7 M_{\odot}$. Point B, the minimum
of $m$ if $t_{evap} >$ 10 Gyr, is now characterized by $m \approx 1.1
\times 10^6 M_{\odot}, a \approx 0.054$ pc, and $M \approx 2.5 \times 10^8
M_{\odot}$. In the latter case, the attainable lower limit on $m$ will be
interestingly close to its upper limit of $M_R=2.6 \times 10^6 M_\odot$.

\section{Summary}

In this paper, we have developed a stochastic model to describe the
dynamics of a black hole near the center of a dense stellar system. The
total force on the black hole is decomposed into a slowly varying part
originating from the response of the whole stellar system, and a random,
rapidly fluctuating part originating from discrete encounters with
individual stars.  We have shown that the time scale over which the latter
force fluctuates is very short compared with the time scale over which the
former changes; hence the justification for the separation of the total
force into these two independent components.  The slowly varying force
itself is approximated as the sum of two contributions: the force on the
black hole due to the potential of the whole stellar system and a force of
dynamical friction which causes the black hole to decelerate as it moves
through the stellar system.

The stochastic force at the position of the black hole is assumed to have a
zero average and to be essentially uncorrelated with itself over time
scales that are short compared with the characteristic period over which
the velocity of the black hole changes considerably, but long enough for
many independent fluctuations of the stochastic force to have occurred.

If the stellar system is approximated by a Plummer model, then the problem
essentially reduces to describing the Brownian motion of a particle in a
harmonic potential. We have shown that after long times, the black hole's
velocity has a zero average and its average location coincides with the
center of the stellar potential.  However, the root mean squared position
and velocity tend towards non-zero values which are independent of time and
of the black hole's initial position and velocity. The steady-state time
autocorrelation functions of the position and velocity were shown to be
approximately given by damped cosine functions. For a Maxwellian
distribution of stars, strict equipartition of kinetic energy between the
black hole and the stars is achieved.

The model was completed by solving for the one unknown parameter --- the
mean squared magnitude of the stochastic force on the black hole --- by
making use of the close relationship between processes of fluctuation and
dissipation, according to which the ``heating'' by the fluctuating force
equals the ``cooling'' caused by the dissipative force of dynamical
friction (the third force on the black hole --- the force due to the
stellar potential --- is conservative).

A Fokker-Planck equation for the diffusion of the probability distribution
of the black hole's position and velocity was developed and solved. The
solution implies that in the steady state these variables are distributed
independently as Gaussians.

The predictions of the model were tested by comparing with the
results of various N-body simulations; the agreement is good, thus
justifying the elements of the model.  In the simulations, we find possible
signs of the existence of weakly damped coherent modes associated with the
stellar system (Weinberg 1994). The total force at the origin of the system
has a strong Fourier component at the frequency of fundamental oscillations
characteristic of the approximately harmonic form of the stellar potential
near the center. If we consider the black hole to be a test particle
affected by such a force, then the autocorrelation functions of the black
hole's position and velocity would be found not to damp with the time
lag, but to persist; the results of simulation do indeed show the presence
of such oscillations at very low amplitude for arbitrarily long time lags.

Finally, we applied the results of the model to Sgr A*. Observational
limits on the peculiar motion of Sgr A* were used to obtain a lower limit
on its mass, under the assumption that it is localized near the center of a
system of equal mass stars distributed according to the Plummer model. More
stringent limits were then deduced by requiring that the evaporation
timescale of the cluster of stars be larger than 1 Gyr.

The Plummer model is a reasonable choice for the black hole problem
considered here since it results in a separable system of linear stochastic
differential equations; it is not clear that arbitrary potential-density
pairs would give equations that are similarly tractable. In particular,
what would happen in the case of black holes at the centers of galaxies if
those are described by singular power-law density profiles? A simple-minded
generalization of our method would not necessarily work (for example,
because of possible divergences in the distribution function).  However, we
believe that our model captures the qualitative features of more
complicated situations for two related reasons.
First, we note that the Plummer model is not an equilibrium solution for a
cluster of stars with a black hole present (see, e.g., Huntley \& Saslaw
1975 and Saslaw 1985); even in non-singular models such as the Plummer
model, the black hole ultimately induces a density cusp.  We have ignored
this complication in our model and found that the model nevertheless
provides a good description of detailed numerical simulations. Second, the
black hole tends to carry its cusp of bound stars with it as it moves
around; thus, it is as if a black hole of a somewhat larger effective mass
were moving in a background consisting of unbound stars. With the cusp
effectively removed, the density profile of this background would be flat
near the center. Since the restoring force and dynamical friction are
provided mainly by the unbound stars, we believe that the essential
components of our model are still valid. 
(For similar reasons -- see \S 5 for details -- the fact that our 
simulations use a fixed stellar potential is not expected to alter 
our conclusions.)
We have carried out numerical
simulations for the case of a particular density profile with a
singularity, namely the Hernquist (1990) model.  We find that the
qualitative results are similar to those described here. In particular, the
early-time autocorrelation functions of the black hole's force and velocity
continue to be described well by damped cosine functions of fixed
frequency. The detailed characterization of the black hole behavior in
terms of the model parameters is different, and requires a more careful
calculation.

\bigskip
\bigskip

We thank G. Quinlan for providing the simulation code, S. Tremaine for
enlightening discussions and a careful reading of the manuscript, and
M.J. Reid and G. Rybicki for useful discussions.  
We also thank the editor, Ethan Vishniac, for his helpful advice.
This work was supported
in part by NASA grants NAG 5-7039, 5-7768, and by NSF grants AST-9900877,
AST-0071019 (for AL).

\begin{figure}[t]
\centerline{\epsfysize=6.0in\epsffile{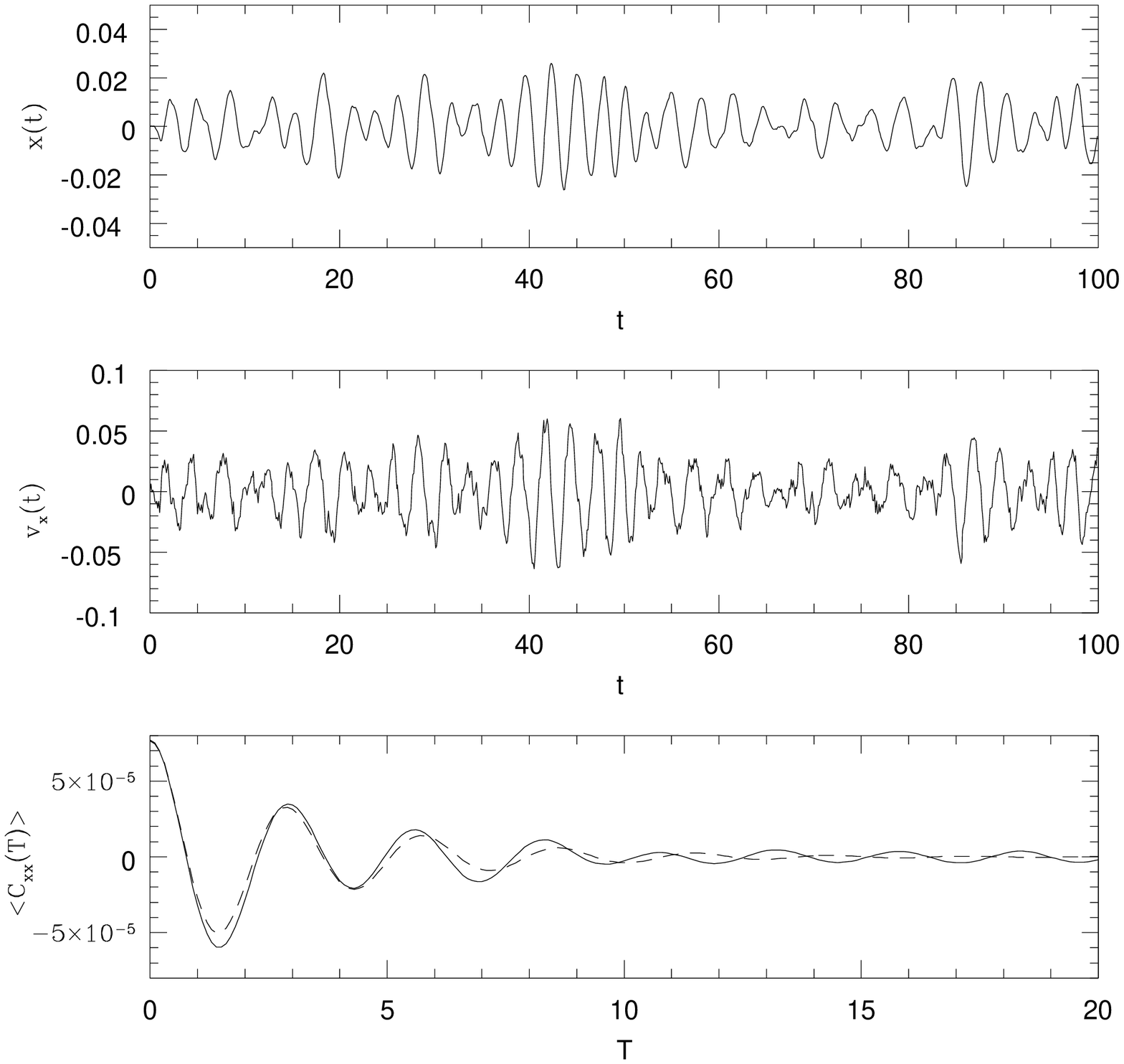}}
\caption{
Results from a simulation of a black hole of mass $m=0.01$ in
a Plummer model of total mass $M=1$ and $N=100,000$ stars. The
evolution of the $x$-components of the black hole's position and
velocity are shown from $t=0$ to $t=100$. The third panel shows the 
autocorrelation function of $x$ for time lags from $T=0$ to
$T=20$. The solid line is the autocorrelation
function obtained from the simulation, and the broken line is that
predicted by our model.
}
\label{fig:1}
\end{figure}

\begin{figure}[t]
\centerline{\epsfysize=6.0in\epsffile{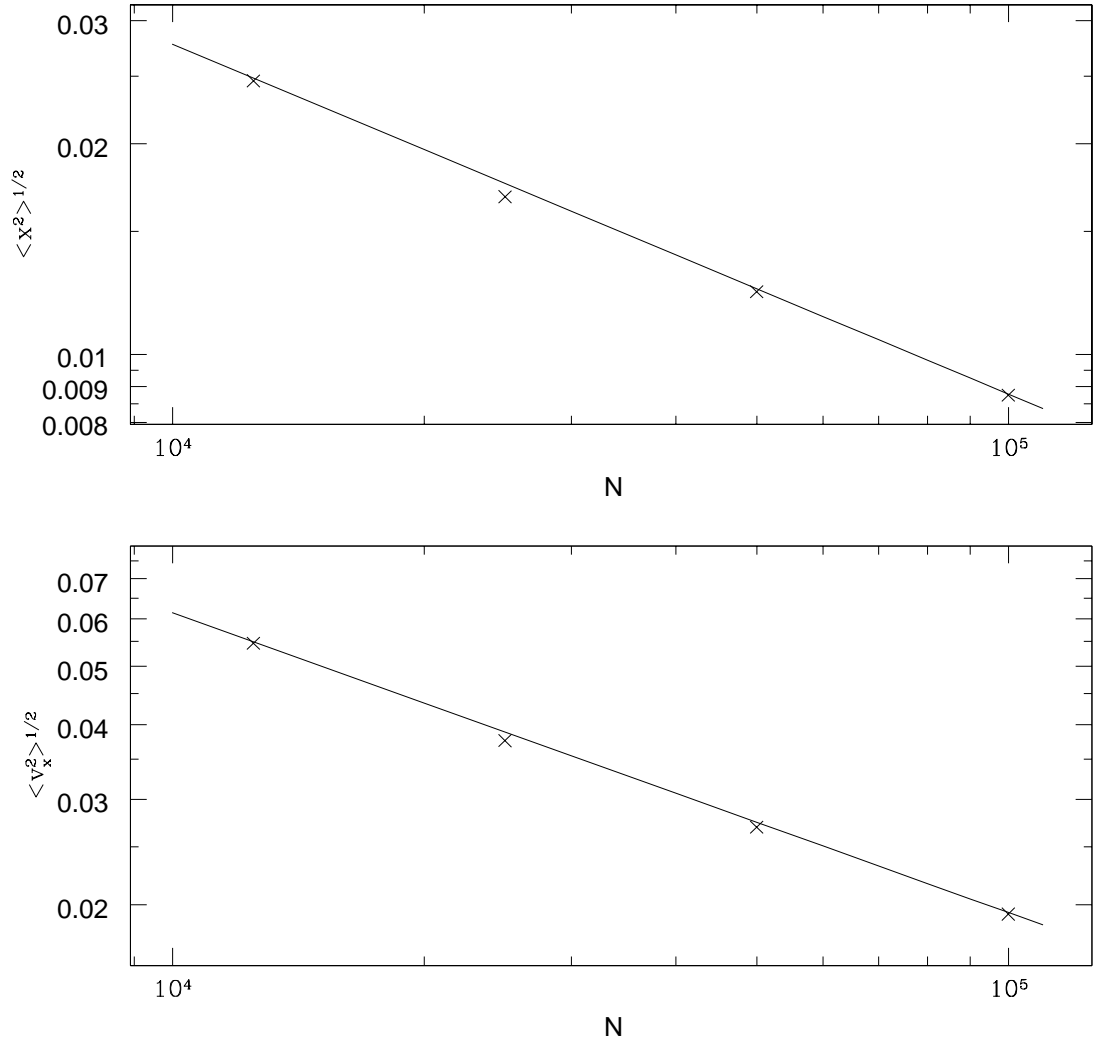}}
\caption{
Results from simulations of a black hole of mass $m=0.01$ in
a Plummer model of total mass $M=1$ and $N=\, $ 12,500, 25,000, 50,000
and 100,000 stars. The panels show the root mean squared values of the 
$x$-components of the black hole's position and velocity. The crosses 
show the results obtained from the
simulations, and the solid lines are the values predicted by our model.  
In each case, the simulation was stopped at time $t=600$.
}
\label{fig:2}
\end{figure}

\begin{figure}[t]
\centerline{\epsfysize=6.0in\epsffile{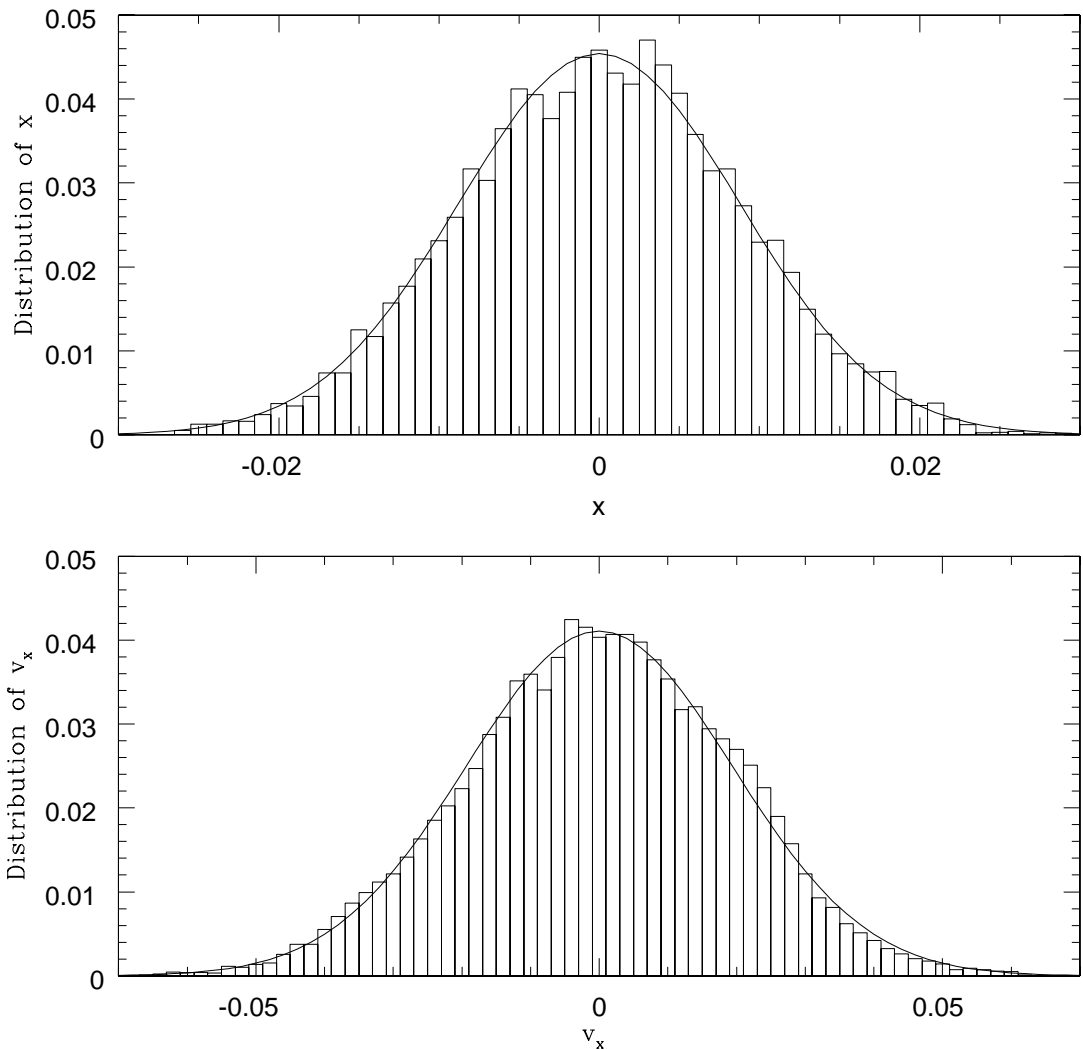}}
\caption{
Results from a simulation of a black hole of mass $m=0.01$ in a Plummer
model of total mass $M=1$ and $N=100,000$ stars. The distributions of the
$x$-components of the black hole's position and velocity are shown as
histograms. The solid lines show the corresponding bin values according to
the analytic model of this paper. 
}
\label{fig:3}
\end{figure}

\begin{figure}[t]
\centerline{\epsfysize=6.0in\epsffile{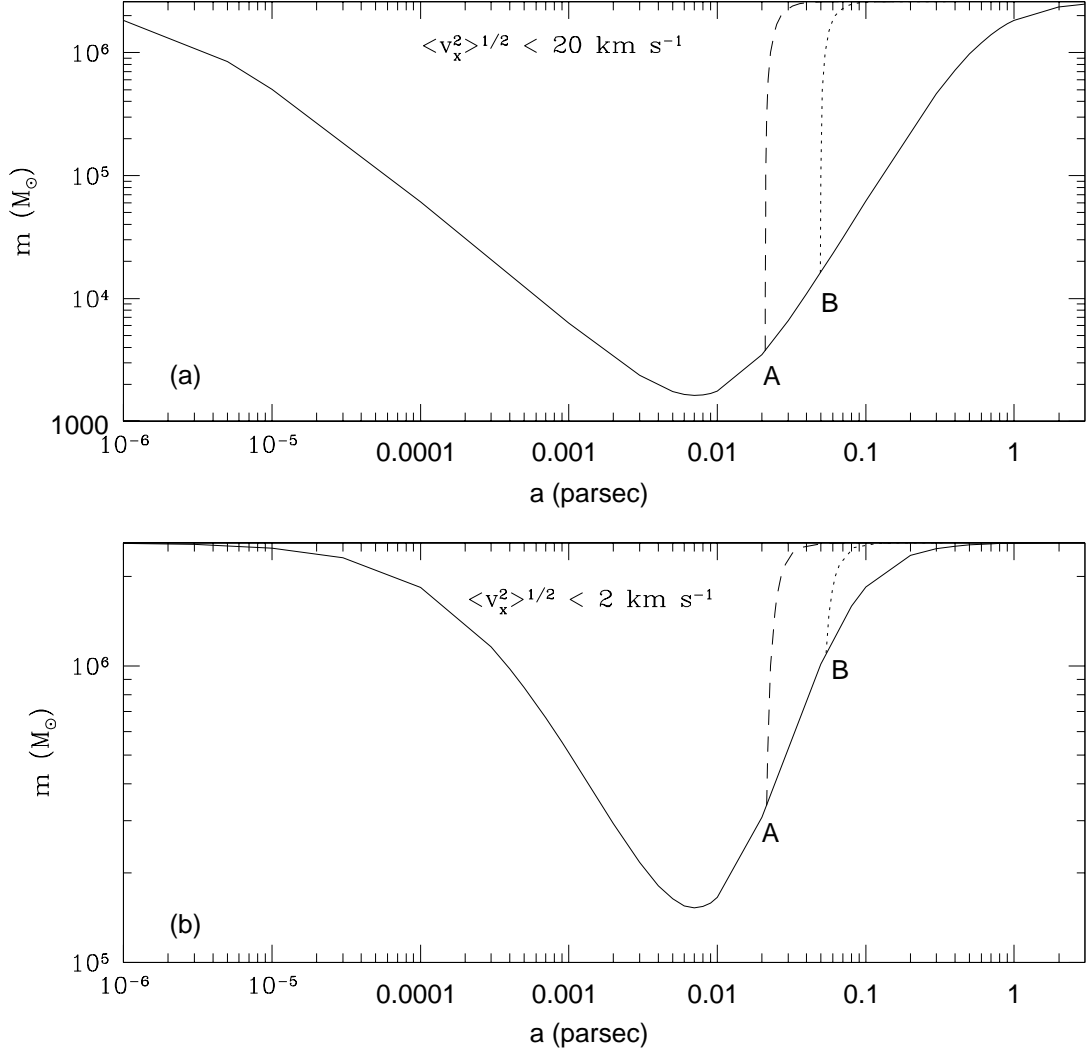}}
\caption{ The region above the solid line in the upper panel denotes the
allowed values of $m$, the mass of the putative black hole in the Galactic
center Sgr A*, and $a$, the length scale characterizing the cluster of
stars (each of mass 1 $M_{\odot}$) surrounding Sgr A*, given that the {\it
rms} value of one component of the black hole's velocity is $<$ 20 km
s$^{-1}$. The region to the right of the dashed line denotes the allowed
values of $m$ and $a$ if the evaporation timescale of the cluster,
$t_{evap}$, exceeds 1 Gyr, and the region to the right of the dotted line
denotes the allowed values if $t_{evap} >$ 10 Gyr.  The lower panel is the
same as the upper panel, except that the {\it rms} value of one component
of the black hole's velocity is now assumed to be $<$ 2 km s$^{-1}$.  }
\label{fig:4}
\end{figure}

\end{document}